\documentclass[pdflatex,sn-mathphys-num]{sn-jnl}

\usepackage{graphicx}%
\usepackage{multirow}%
\usepackage{amsmath,amssymb,amsfonts}%
\usepackage{amsthm}%
\usepackage{mathrsfs}%
\usepackage[title]{appendix}%
\usepackage{xcolor}%
\usepackage{textcomp}%
\usepackage{manyfoot}%
\usepackage{booktabs}%
\usepackage{algorithm}%
\usepackage{algorithmicx}%
\usepackage{algpseudocode}%
\usepackage{listings}%

\theoremstyle{thmstyleone}%

\theoremstyle{thmstyletwo}%

\theoremstyle{thmstylethree}%

\begin{document}

\title[Comments on the non-existence of unified hoop conjecture]{Comments on the non-existence of unified hoop conjecture}

\author[1]{\fnm{Amrita} \sur{Bhattacharya}}\email{bamrita323@gmail.com}
\equalcont{These authors contributed equally to this work.}

\author*[2]{\fnm{Ramil N.} \sur{Izmailov}}\email{izmailov.ramil@gmail.com}

\author[2]{\fnm{Ramis Kh.} \sur{Karimov}}\email{karimov\_ramis\_92@mail.ru}
\equalcont{These authors contributed equally to this work.}

\affil[1]{\orgdiv{Department of Mathematics}, \orgname{Kidderpore College}, \orgaddress{\street{2, Pitamber Sircar Lane}, \city{Kolkata}, \postcode{700023}, \state{WB}, \country{India}}}

\affil*[2]{\orgdiv{Zel'dovich International Center for Astrophysics}, \orgname{M. Akmullah Bashkir State Pedagogical University}, \orgaddress{\street{3A, October Revolution Street}, \city{Ufa}, \postcode{450008}, \state{RB}, \country{Russia}}}

\abstract{In a recent article, Hod \cite{Hod:2020} has concluded the non-existence, heretofore unnoticed, of a unified Thorne's hoop conjecture that holds for the quasi-local mass, $\mathcal{M}(r\leq R_{+})$ for static Reissner-Nordstr\H{o}m black hole, but does not hold for the spinning electrically charged Kerr-Newman black holes, when the same quasi-local mass is used. While the conclusion is correct and important, we wish to point out a curious exception for which the conjecture holds in a unified way for both the static and spinning \textit{tidal} charged $4d$ braneworld black holes of string theory, when the mass in the conjecture is the same quasi-local mass.}

\maketitle

The famous hoop conjecture of Kip Thorne \cite{Thorne:1972} describes necessary and sufficient condition for the formation of black hole horizons in the dynamical gravitational collapse. There are several equivalent versions of the hoop conjecture. The simplest, and original, one states that an imploding object of mass $\mathcal{M}$ forms a black hole when, and only when, a circular hoop of critical circumference $\mathcal{C}=4\pi \mathcal{M}$ could be placed around it and rotated 360$^{0}$ in all azimuthal directions to form an engulfing sphere. The hoop conjecture for black holes therefore can be stated in a simple form \cite{Hod:2020}
\begin{equation}
\mathcal{H}\equiv \frac{4\pi \mathcal{M}}{\mathcal{C}}\geq 1 \; \Rightarrow \; \text{black hole horizon exists}.
\end{equation}

Thorne \cite{Thorne:1972} did not specify the mass $\mathcal{M}$. If it is regarded as the asymptotic ADM mass $M_{\infty}$ then the conjecture appears to hold even for a \textit{horizonless} charged compact mass thereby weakening the conjecture \cite{deLeon:1987, Bonnor:1983}. To avoid this scenario, Hod \cite{Hod:2018} earlier proposed that the mass in the conjecture should be interpreted as the quasilocal mass $\mathcal{M}$ contained within the horizon radius and not the asymptotic ADM mass $M$. We shall see that this interpretation satisfies the conjecture (1) for braneworld black holes of string theory.

A brief sketch of the braneworld black holes is in order. Recent developments in string theory have shown that matter fields can be localized on a $3-$brane in $1+3+d$ dimensions, while gravity can freely propagate in the $d$ extra dimensions beyond the four \cite{Shiromizu:2000}. In the Randall-Sundrum scenario \cite{Randall:1999} black holes can be localized on a $3-$brane where gravity propagates in the $5d-$ bulk. It was shown by Dadhich, Maartens, Papadopoulos and Rezania (DMPR) \cite{Dadhich:2000} that the Reissner-Nordstr\H{o}m (RN) metric is an exact solution of the closed system of effective $4d-$Einstein field equations on the brane with the source term modified by the projection of bulk effects on the brane. The solution is re-interpreted as a black hole without the electric charge, but with instead a tidal charge arising via gravitational effects from the projection. The solution could in principle arise in the strong-gravity regime of black holes \cite{Dadhich:2000}. The strong lensing characteristics observed in recent Event Horizon Telescope (EHT) measurements can constrain $q$ as has been recently found in \cite{Nandi:2024}.

The $5d$ field equations of \cite{Shiromizu:2000}, when induced on the brane is the standard $4d-$Einstein equations of general relativity with new source terms on the brane, which leads to the vacuum field equations \cite{Dadhich:2000}:
\begin{equation}
R_{\mu \nu }=-\mathcal{E}_{\mu \nu }, \quad R_{\mu }^{\mu} = 0 = \mathcal{E}_{\mu }^{\mu }
\end{equation}%
where $\mathcal{E}_{\mu \nu }$ is the limit on the brane of the projected $5d-$ bulk Weyl tensor $\mathcal{E}_{AB} = - C_{ACBD} n^{C}n^{D}$. ($A, B = 0 - 4; \mu, \nu = 0 - 3)$. Because of the symmetries of the Weyl tensor, $\mathcal{E}_{AB}$ is symmetric and tracefree. Also it has no components orthogonal to the brane so that $\mathcal{E}_{AB}\rightarrow \mathcal{E}_{\mu \nu}\delta _{A}^{\mu }\delta _{B}^{\nu }$ on the brane.

In view of the Bianchi identities on the brane, the integrability condition is 
\begin{equation}
\nabla ^{\mu }\mathcal{E}_{\mu }^{\nu } = 0
\end{equation}%
where $\nabla _{\mu }$ is the brane covariant derivative. Further details are to be found in \cite{Dadhich:2000}. 

One branch of the braneworld solution of Eq.(2) is just the Reissner-Nordstr\H{o}m-like metric with $+q^{2}$ now as the tidal charge correction to Schwarzschld potential and the other solution is the braneworld black hole metric \cite{Dadhich:2000} with a \textit{negative} correction $-q^{2}$  ($q^{2}>0$) so that the metric becomes
\begin{equation}
d\tau ^{2}=-\left( 1-\frac{2M}{R}-\frac{q^{2}}{R^{2}}\right) dt^{2}+\left( 1-%
\frac{2M}{R}-\frac{q^{2}}{R^{2}}\right) ^{-1}dR^{2}+R^{2}d\Omega ^{2}
\end{equation}%
where $d\Omega^{2}$ is the metric on a unit sphere. The negative tidal correction $-q^{2}$ can be formally achieved by the transcription: electric charge $Q\rightarrow $ $iq$ in the RN metric. There is now only a single spherical horizon, which is bigger than the Reissner-Nordstr\H{o}m horizon and is given by
\begin{equation}
R_{+}=M+\sqrt{M^{2}+q^{2}}.
\end{equation}%
Fot this static Reissner-Nordstr\H{o}m\ case with the above transcription, hoop conjecture trivially holds, with $\mathcal{C=}2\pi R_{+},$ 
\begin{equation}
\mathcal{M}=M+\frac{q^{2}}{2R_{+}}\Rightarrow \frac{4\pi \mathcal{M}}{\mathcal{C}}=1.
\end{equation}%
which is just the definition of the horizon or the null surface $g_{00}(R_{+})=0$.

The negative correction to the Schwarzschild potential on the brane is the key that ensures the existence of a unified hoop conjecture in contrast to what happens in the Kerr-Newman case. The spinning tidal charged black hole is the Kerr-Newman metric with the transcription $Q\rightarrow iq$.

The spinning tidal charged braneworld black hole is given by \cite{Aliev:2005}
\begin{equation}
d\tau^{2} = -\frac{\Delta}{\Sigma} \left(dt - a\sin^{2}{\theta} d\phi\right)^{2} + \Sigma \left(\frac{dR^{2}}{\Delta} + d\theta^{2}\right) + \frac{\sin^{2}{\theta}}{\Sigma} \left[a dt - \left(R^{2} + a^{2}\right) d\phi\right]^{2},
\end{equation}
with
$$\Delta = R^{2} - 2MR + a^{2} - q^{2}, \quad \Sigma = R^{2} + a^{2} \cos^{2}{\theta},$$
where $a$ is the spin parameter (angular momentum per unit mass, $a=J/M$). The metric is the spinning generalization of the static metric (4). The event horizon is determined by largest root of the equation $\Delta =0$, given by
\begin{equation}
R_{+} = M + \sqrt{M^{2} - a^{2} + q^{2}}.
\end{equation}
Using the metric (8), the equatorial circumference of the black hole (7) can
be obtained as 
\begin{equation}
\mathcal{C}=4\pi M\left[ 1+\frac{q^{2}}{2M\left( M+\sqrt{M^{2}+a^{2}-q^{2}}%
\right) }\right].
\end{equation}

Following exactly the same arguments as that of Hod \cite{Hod:2020}, one now has the interior mass $\mathcal{M}$ enclosed by the horizon $R=R_{+}$ as
\begin{equation}
\mathcal{M}(R=R_{+})=M+\frac{q^{2}}{4R_{+}}\left[1+\frac{R_{+}^{2}+a^{2}}{%
aR_{+}}\arctan \left( \frac{a}{R_{+}}\right)\right] .
\end{equation}%
Exactly following Hod \cite{Hod:2020}, we easily find that the conjecture%
\begin{equation}
\mathcal{H\equiv }\frac{4\pi \mathcal{M}}{\mathcal{C}}>1
\end{equation}%
holds for the tidal charged spinning braneworld black hole.

\begin{table}[!ht]
\centering
\begin{tabular}{ccccccc}
\hline
$\mathcal{H}$ & $a=0$ & $a=0.2M$ & $a=0.4M$ & $a=0.6M$ & $a=0.8M$ & $a=M$ \\ 
\hline
$q=0$ & $1.00000$ & $1.00000$ & $1.00000$ & $1.00000$ & $1.00000$ & $1.00000$
\\ 
$q=0.2M$ & $1.00000$ & $1.00003$ & 1.00014 & 1.00038 & 1.00093 & 1.00338 \\ 
$q=0.4M$ & $1.00000$ & $1.00012$ & 1.00051 & 1.00133 & 1.00307 & 1.00842 \\ 
$q=0.6M$ & $1.00000$ & 1.00022 & 1.00095 & 1.00243 & 1.00535 & 1.01228 \\ 
$q=0.8M$ & $1.00000$ & 1.00032 & 1.00136 & 1.0034  & 1.00714 & 1.01468 \\ 
$q=M$ & $1.00000$ & 1.0004 & 1.00167 & 1.0041 & 1.00831 & 1.01591 \\ \hline
\end{tabular}
\caption{Specific values of the dimensionless ratio $\mathcal{H} = \mathcal{H}(a,q)$ for various values of the black-hole parameters $a$ and $Q$.}
\end{table}

Table 1 presents the values of the dimensionless function $\mathcal{H}(a,q)$ for different values of the angular momentum $a$ and charge $q$. Table shows that for a fixed angular momentum $a$, the value of $\mathcal{H}(a,q)$ increases monotonically and reaches its maximum value at $q=M$. For a fixed $q$, the values of $\mathcal{H}(a,q)$ also increase monotonically. Thus, for a braneworld black hole, the function $\mathcal{H}(a,q)\geq 1$ for any values of $a$ and $q$.

\begin{figure}[!ht]
  \centerline{\includegraphics[scale=0.4]{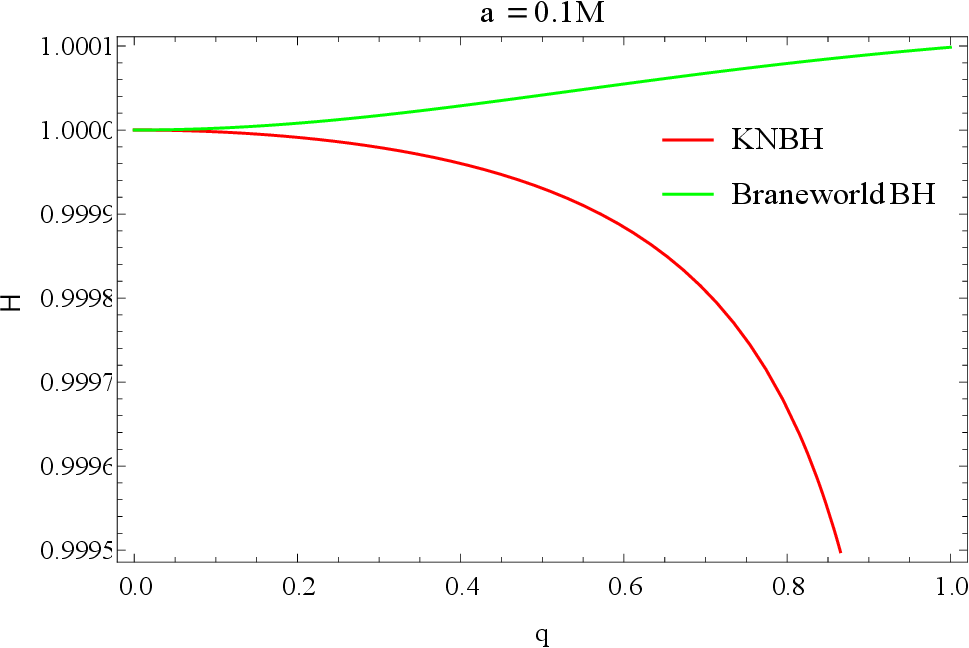} \includegraphics[scale=0.4]{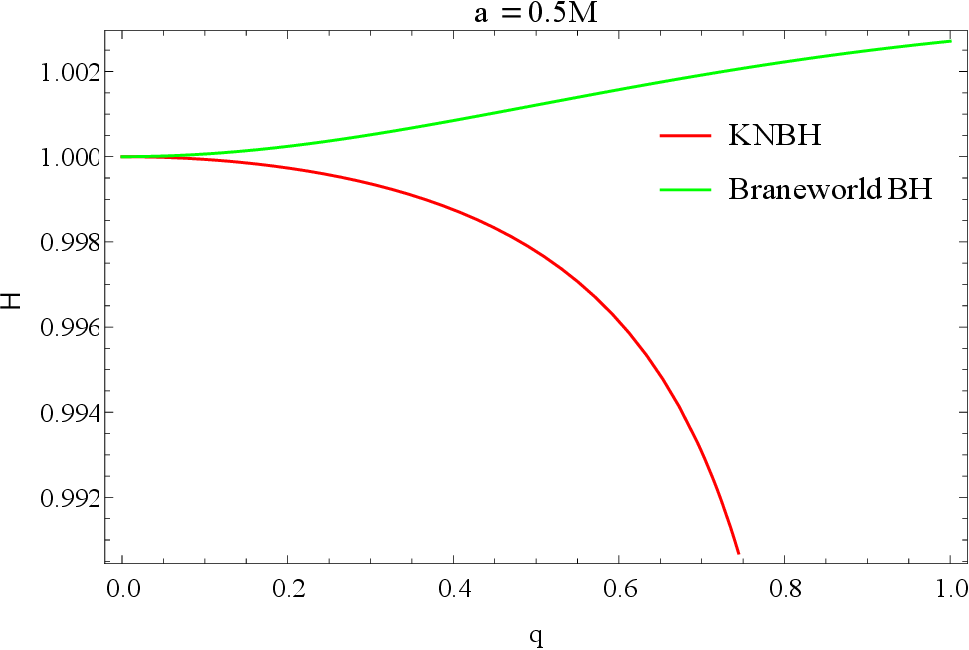}}
  \caption{Plots of dimensionless functions $\mathcal{H}(a,q)$ as a function of charge for $a=0.1M$ and $a=0.5M$.}
  \label{fig3}
\end{figure}

Figure 1 shows the plots of the dimensionless functions $\mathcal{H}(a,q)$ of Kerr-Newman black hole (KNBH) and braneworld black hole (Branewold BH) as a function of charge for $a=0.1M$ and $a=0.5M$. From the Figure 1 it is clear that for the braneworld black hole the dimensionless function is $\mathcal{H}(a,q) \geq 1$, and for the Kerr-Newman black hole $\mathcal{H}(a,q) \leq 1$.

Defining the energy-momentum components as $T^{00}=\rho $, $T^{11}=-p_{R}$, $T^{22}=-p_{\theta}$ and $T^{33}=-p_{\phi}$, the energy conditions are given in the form \cite{Visser:1996}
\begin{eqnarray}
\text{NEC}&:& \rho + p_{i} \geq 0, \quad i = R, \theta, \phi, \\
\text{WEC}&:& \rho \geq 0, \quad \rho + p_{i} \geq 0, \quad i = R, \theta, \phi, \\
\text{SEC}&:& \rho + \sum p_{i} \geq 0, \quad i = R, \theta, \phi,
\end{eqnarray}
where null energy condition denoted as NEC, weak energy condition as WEC and strong energy condition as SEC.

The non-zero components of the energy-momentum tensor  of braneworld black hole are as follows \cite{Azreg:2014}
\begin{equation}
\rho = -p_{R} = p_{\theta} = p_{\phi} = - \frac{q^{2}}{\left(R^{2} + a^{2} \cos^{2}{\theta}\right)^{2}}.
\end{equation}

Substituting Eq.(14) into Eqs.(11)-(13) we get
\begin{eqnarray}
\textmd{NEC}&:& \rho + p_{R} = 0, \quad \rho + p_{\theta} = \rho + p_{\phi}
= - \frac{2 q^{2}}{\left(R^{2} + a^{2} \cos^{2}{\theta}\right)^{2}}, \\
\textmd{WEC}&:& \rho = - \frac{q^{2}}{\left(R^{2} + a^{2} \cos^{2}{\theta}%
\right)^{2}}, \quad \rho + p_{R} = 0,  \nonumber \\
&&\rho + p_{\theta} = \rho + p_{\phi} = - \frac{2 q^{2}}{\left(R^{2} + a^{2}
\cos^{2}{\theta}\right)^{2}}, \\
\textmd{SEC}&:& \rho + p_{R} + p_{\theta} + p_{\phi} = - \frac{2 q^{2}}{%
\left(R^{2} + a^{2} \cos^{2}{\theta}\right)^{2}}.
\end{eqnarray}

From equations (15), (16) and (17) we can conclude that the NEC, WEC and SEC are violated, since they always take a negative value.

In conclusion, we state that Hod \cite{Hod:2020} has indeed brought to light a genuine problem of non-existence of a unified hoop conjecture in connection with Kerr-Newman black holes. Here we pointed out an intriguing fact that introducing a kind of Wick rotation $Q\rightarrow $ $iq$ in the Kerr-Newman metric converts it into a tidal charge braneworld metric that can yield a unified hoop conjecture but the price is that we are no longer dealing with the familiar electric charge.

\end{document}